\documentclass[aps,print,onecolumn,hidepacs,hidekeys]{revtex4}%
\usepackage{amssymb}
\usepackage{amsfonts}
\usepackage{amsmath}
\usepackage{graphicx}%
\usepackage{xcolor}
\providecommand{\U}[1]{\protect\rule{.1in}{.1in}}

\begin{document}
\title{Hybrid Airy Plasmons with Dynamically Steerable Trajectories}
\author{Rujiang Li,\textit{$^{a}$} Muhammad Imran,\textit{$^{b}$} Xiao Lin,\textit{$^{a}$}
 Huaping Wang,\textit{$^{b}$} Zhiwei Xu,\textit{$^{b}$} and Hongsheng Chen\textit{$^{a}$}}
\email{hansomchen@zju.edu.cn}
\affiliation{$^{a}$State Key Laboratory of Modern Optical Instrumentation, Zhejiang
University, Hangzhou 310027, China.\\
$^{b}$Institute of Marine Electronics Engineering, Zhejiang University, Hangzhou 310058, China.}

\begin{abstract}
With the intriguing
properties of diffraction-free, self-accelerating,
and self-healing, Airy plasmons are promising to be used in
the trapping, transporting, and sorting of micro-objects, imaging,
and chip scale signal processing.
However, the high dissipative loss and the lack of dynamical
steerability restrict the implementation of Airy plasmons in these applications.
Here we reveal the hybrid Airy plasmons for the first time
by taking a hybrid graphene-based plasmonic waveguide in the terahertz (THz) domain
as an example.
Due to the coupling between an optical mode and a plasmonic mode,
the hybrid Airy plasmons can have large
propagation lengths and effective transverse deflections,
where the transverse waveguide confinements are governed by the hybrid
modes with moderate quality factors.
Meanwhile, the propagation trajectories of hybrid Airy plasmons
are dynamically steerable by changing the chemical potential of graphene.
These hybrid Airy plasmons may promote the further discovery of non-diffracting
beams with the emerging developments of optical tweezers and tractor beams.

\end{abstract}
\maketitle



\section{Introduction}

With the analogies between Schr\"{o}dinger equation and paraxial wave equation,
the concept of Airy beams is extended from quantum
mechanics to optics to describe a kind of non-diffracting beams
\cite{AJP,OL32-979,LPR4-529}, where the field amplitude is truncated to ensure
the containment of finite energy \cite{OL32-2447} and to enable the
experimental realization \cite{PRL99-213901}. Due to the intriguing properties
of diffraction-free, self-accelerating \cite{OL32-979}, self-healing
\cite{OE16-12880}, and abruptly autofocusing \cite{OL35-4045,OL36-1842}, Airy
beams are promising in a serials of applications, including the trapping,
transporting, and sorting of micro-objects \cite{nphoton2-675,
OL36-2883,AO50-43}, imaging \cite{OL40-5686}, and chip scale signal processing
\cite{APL102-101101,OL39-5997}. Besides, the concept of Airy beams has been
extended into various fields of physics, such as temporal pulses
\cite{nphoton4-103,OE16-10303,OE19-2286}, spin waves \cite{PRL104-197203},
water waves \cite{PRL115-034501}, and matter waves \cite{nature,PRA87-043637}.

Considering potential applications in flatland devices, one-dimensional
Airy beams propagating on metal-dielectric or graphene-dielectric interfaces
in the form of surface plasmon polaritons with subwavelength transverse waveguide confinements
were theoretically proposed \cite{OL35-2082,ieeepj} and experimentally realized
\cite{PRL107-116802,PRL107-126804}. However, the propagation lengths of Airy
plasmons are usually short owing to the strong Ohmic losses in metal and graphene,
which brings a challenge to the \textcolor[rgb]{0.00,0.00,0.00}{observation} and implementation
of Airy plasmons \textcolor[rgb]{0.00,0.00,0.00}{experimentally} in the framework
of paraxial wave equation \cite{OL34-3430}.
Meanwhile, the traditional methods to steer beam propagation
trajectories such as tuning the excitation source \cite{OL36-3191} and
creating linear optical potentials \cite{OL36-1164,OL38-1443} are either
cumbersome or non-real time,
although the dynamical steerability is essential in optical
micromanipulation and signal processing. Thus, searching for a better platform
to realize the low-loss and dynamically steerable Airy plasmons
is highly imperative \cite{LPR8-221}.

In this paper, inspired by the hybrid plasmonic waveguides
where both subwavelength confinement and long range propagation are fulfilled \cite{nphoton2-496,JLT32-3597},
we reveal the hybrid Airy plasmons for the first time by taking
a hybrid graphene-based plasmonic waveguide in the THz domain as an example.
Due to the coupling between an optical mode and a plasmonic mode,
the hybrid Airy plasmons can have large
propagation lengths and effective transverse deflections, where the transverse waveguide
confinements are governed by the hybrid modes with moderate quality factors.
Meanwhile, since the chemical potential of graphene can be tuned by applying a gate voltage,
the propagation trajectories of hybrid Airy plasmons are dynamically steerable.

\section{Results and discussion}
\subsection{Model equation}

For the quasi-TM Airy plasmons propagating in a planar plasmonic
waveguide, the governing paraxial wave equation for the amplitude $\psi$ is%
\begin{equation}
i\frac{\partial\psi\left(  s,\xi\right)  }{\partial\xi}+\frac{\partial^{2}%
\psi\left(  s,\xi\right)  }{\partial s^{2}}=0, \label{schrodinger}%
\end{equation}
where $\psi$ is related with the magnetic field $H_{y}$,
$s=y/y_{0}$ is the dimensionless coordinate in the transverse direction that
is parallel to the interfaces, $\xi=z/2\beta
y_{0}^{2}$ is the dimensionless complex propagation distance,
$\beta=\beta_{r}+i\beta_{i}$
is the propagation constant of the waveguide mode distributed
in $x$ direction with $\beta_{r}\equiv\text{Re}\left(\beta\right)$ and
$\beta_{i}\equiv\text{Im}\left(\beta\right)$,
and $y_{0}$ is an arbitrary transverse scale. The solution of
Airy plasmons with finite energy at the input of $\psi\left(  s,0\right)
=\text{Ai}\left(  s\right)  \exp\left(  as\right)  $ is%
\begin{equation}
\psi\left(  s,\xi\right)  =\text{Ai}\left(  s-\xi^{2}+i2a\xi\right)
\exp\left[  i\left(  s\xi+a^{2}\xi-\frac{2}{3}\xi^{3}\right)  \right]
\exp\left(  as-2a\xi^{2}\right)  , \label{solution}%
\end{equation}
where $a$ is a positive decay factor to truncate the amplitude at the negative
infinity, and the width of main lobe of Airy plasmons
is approximated by $2y_{0}$ \cite{OL35-2082}.
According to the integral representation of Airy function
\cite{Airy-function}, Eq. (\ref{solution}) can also be built using plane
waves
\begin{equation}
\psi\left(  s,\xi\right)  =\frac{1}{2\pi}\int_{-\infty}^{+\infty}\Phi\left(
k_{s},\xi\right)  \exp\left(  ik_{s}s\right)  dk_{s}, \label{plane wave}%
\end{equation}
where
\begin{equation}
\Phi\left(  k_{s},\xi\right)  =\exp\left(  \frac{a^{3}}{3}\right)  \exp\left(
-ia^{2}k_{s}\right)  \exp\left(  -ak_{s}^{2}\right)  \exp\left(  i\frac
{k_{s}^{3}}{3}\right)  \exp\left(  -ik_{s}^{2}\xi\right)  \label{Fourier}%
\end{equation}
is the Fourier spectrum in $k$-space, the cubic phase term $\exp\left(
ik_{s}^{3}/3\right)  $ is associated with the spectrum of Airy plasmons, the first
Gaussian function $\exp\left(  -ak_{s}^{2}\right)  $ arises from the
exponential apodization of the beam, and the second Gaussian function
$\exp\left(  k_{s}^{2}\ \xi_{i}\right)  $ is originated from the last term in Eq. (\ref{Fourier})
with
$\xi_{i}\equiv \text{Im}\left( \xi \right)=-\beta_{i}z/\left[  2\left(
\beta_{r}^{2}+\beta_{i}^{2}\right)  y_{0}^{2}\right]$.
From Eqs. (\ref{plane wave})-(\ref{Fourier}), the $y$ and $z$ components of
the wavevector are $k_{y}=k_{s}/y_{0}$ and
$k_{z}=\left( \beta_{r}+\delta\beta_{r}\right)
+i\left(\beta_{i}+\delta\beta_{i}\right)  $,
respectively, where $\delta\beta_{r}=-k_{y}^{2}\beta_{r}/\left[  2\left(
\beta_{r}^{2}+\beta_{i}^{2}\right)  \right]  $ and $\delta\beta_{i}=k_{y}%
^{2}\beta_{i}/\left[  2\left(  \beta_{r}^{2}+\beta_{i}^{2}\right)  \right]  $.
To insure the validity of the quasi-TM condition, the
wavevector components must satisfy $\left\vert k_{y}\right\vert \ll\beta_{r}$,
$\left\vert \delta\beta_{r}\right\vert \ll\beta_{r}$, $\left\vert
\delta\beta_{i}\right\vert \ll\beta_{i}$, and the paraxial approximation
$\left\vert \partial^{2}\psi/\partial
z^{2}\right\vert \ll\left\vert 2i\beta\partial\psi/\partial z\right\vert $.
Given the Gaussian spectrum of Airy plasmons in Eq. (\ref{Fourier}), the four
conditions reduce to%
\begin{equation}
\sqrt{a}\beta_{r}y_{0}\gg1. \label{condition_1}%
\end{equation}
Besides, according to Eq. (\ref{solution}), the parabolic self-deflection
experienced by Airy plasmons during propagation can be estimated as
\begin{equation}
y=\frac{z^{2}}{4\beta_{r}^{2}y_{0}^{3}}. \label{parabolic}%
\end{equation}
Taking $L_{a}=1/2\beta_{i}$ as the analytically estimated propagation length
of Airy plasmons \cite{OL35-2082}, the transverse displacement at the
propagation length can be calculated analytically as
\begin{equation}
\Delta y_{a}(z=L_{a})=\frac{1}{16\beta_{r}^{2}\beta_{i}^{2}y_{0}^{3}}.
\label{displacement}%
\end{equation}

From Eq. (\ref{solution}), the decay factor $a$ imposes an attenuation to the
propagation of Airy plasmons, and $\xi_{i}$ introduces extra exponential
terms. These induce errors to the analytical results $L_{a}$ and $\Delta
y_{a}$. Thus we also need to calculate the propagation length and transverse
displacement numerically, and compare them with the analytical results.
For simplicity, the numerical propagation length $L_{n}$ is
defined as the distance where the power $P=\int_{-\infty}^{+\infty
}\left\vert H_{y}\left(  x,y,z\right)  \right\vert ^{2}dy$ decreases to
$e^{-1}P_{0}$ along the propagation direction, where $P_{0}=\int_{-\infty
}^{+\infty}\left\vert H_{y}\left(  x,y,z=0\right)  \right\vert ^{2}dy$ is the
input power at $z=0$. Accordingly, the numerical transverse displacement $\Delta y_{n}$
can be defined as the displacement of the maximum field intensity
from $z=0$ to the propagation length.
Since the solution of Airy plasmons is assumed to be a
perturbation of the waveguide mode, our model is valid if the analytical
propagation length $L_{a}$ and transverse displacement $\Delta y_{a}$ are
approximately equal to the numerical propagation length $L_{n}$ and transverse
displacement $\Delta y_{n}$, respectively.

To realize one-dimensional Airy plasmons in a planar plasmonic waveguide,
Eq. (\ref{condition_1}) must be fulfilled. Since
the decay factor $a$ is usually small to avoid excessively changing the
non-diffracting behavior of Airy plasmons, the real part of propagation
constant $\beta_{r}$ and the transverse scale $y_{0}$ must be large enough.
However, from Eq. (\ref{displacement}) the decrease of the transverse
displacement $\Delta y_{a}$ would be a challenge for the detection and
measurement of Airy plasmons experimentally. A contradiction exists between
the validity of Eq. (\ref{condition_1}) and a large enough transverse
displacement. To solve this problem, plasmonic waveguides with high
quality factors defined as $Q=\beta_{r}/\beta_{i}$ \cite{acsphotonics3-737}
should be used.

\subsection{Hybrid modes}

\begin{figure}[ptb]
\centering
\vspace{-0.0cm} \includegraphics[width=8.5cm]{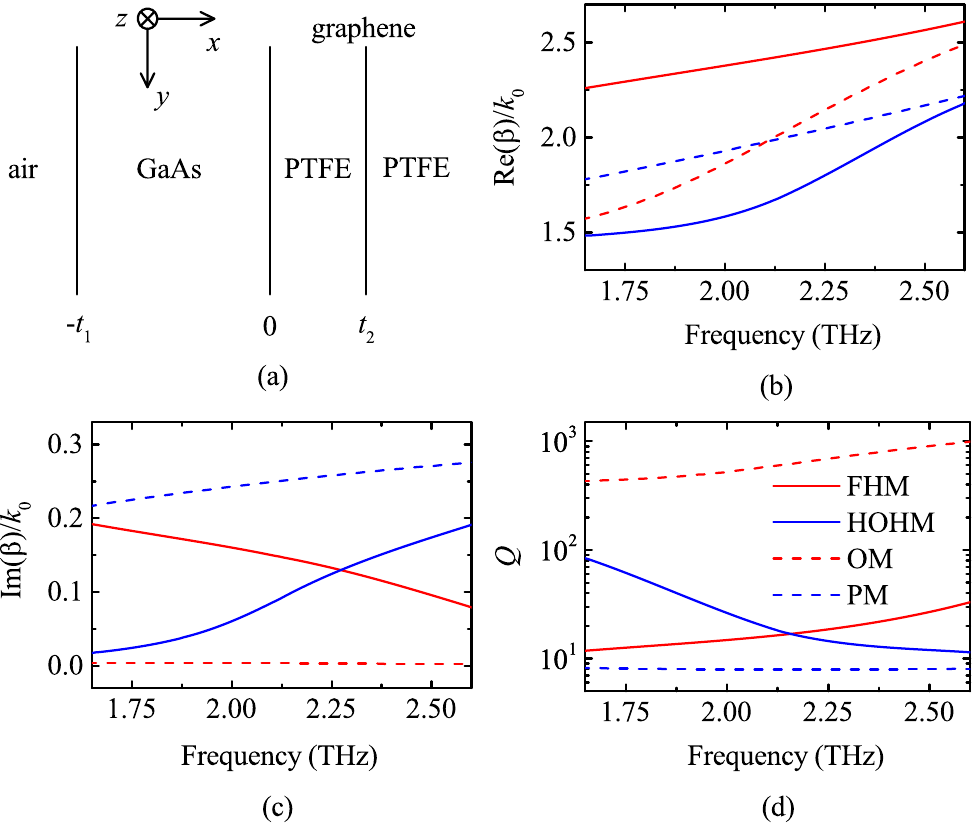} \vspace{0.0cm}%
\caption{(Color Online) (a) Schematic of the hybrid graphene-based planar plasmonic
waveguide, where $z$ is the propagation direction, $x$ is perpendicular to the interfaces,
the thickness of GaAs is $t_{1}=20$ $\mu$m, the thickness of the left PTFE is $t_{2}=10$ $\mu$m,
and the permittivities for air, GaAs, and PTFE are $\varepsilon_{1}=1$,
$\varepsilon_{2}=13.06+0.01i$, and $\varepsilon_{3}=2.08+0.01i$, respectively.
The parameters for the trilayer graphene are $\mu=$ 10 000 $\text{cm}^{2}/(\text{V}\cdot\text{s})$, $T=300$ K,
and $\mu_{c}=0.3$ eV. (b)-(d) The real and imaginary parts of the propagation
constants, and the quality factors of the hybrid modes versus frequency, respectively, where the solid
red curves denote the fundamental hybrid mode (FHM) and the solid blue curves denote the
higher order hybrid mode (HOHM). For comparison, the corresponding curves for the optical mode (OM)
supported by air-GaAs-PTFE and plasmonic mode (PM) supported by PTFE-graphene-PTFE
are also plotted in dashed red and blue, respectively.}
\label{structure}%
\end{figure}

Plasmonic waveguides based on nobel metals usually have low quality factors due to the strong Ohmic
losses in metals. Even for graphene-based plasmonic waveguides, graphene plasmons
also suffer from high dissipative loss \cite{Nanotechnology,IEEEJSQE}, although the surface conductivity of
graphene is almost purely imaginary in the THz domain \cite{science332-1291}.
In contrary, dielectric
optical waveguides have high quality factors since the dielectric loss is
low. If the optical mode in a dielectric optical waveguide and the plasmonic
mode in a plasmonic waveguide are coupled, hybrid
modes with moderate quality factors may exist.

Inspired by the hybrid plasmonic waveguides where both subwavelength confinement and
long range propagation are fulfilled \cite{nphoton2-496,JLT32-3597},
and the flourishing developments of THz science and technology \cite{nmat1-26},
we propose a planar hybrid graphene-based plasmonic waveguide in the THz domain
as an example.
The structure is shown in Fig. \ref{structure}(a),
where $z$ is the propagation direction, $x$ is perpendicular to the interfaces,
the thicknesses of GaAs and the left polytetrafluoroethylene (PTFE) are
$t_{1}=20$ $\mu$m and $t_{2}=10$ $\mu$m, respectively.
The thickness of the right PTFE can be treated as infinite as long as
it is large enough compared with the skin depth of the waveguide mode.
The permittivities for
air, GaAs, and PTFE are $\varepsilon_{1}=1$,
$\varepsilon_{2}=13.06+0.01i$ \cite{optical constants},
and $\varepsilon_{3}=2.08+0.01i$ \cite{JKPS}, respectively.
GaAs and PTFE are widely used in THz waveguides and their permittivities
are both assumed to be constants
since their chromatic dispersions are small in the THz domain.
The surface conductivity of trilayer graphene is $\sigma=3\sigma_{g}$
\cite{NL7-2711,nnano7-330,LPR8-291},
where $\sigma_g$ is the surface conductivity of monolayer graphene calculated
by Kubo formula
with $\mu=$ 10 000 $\text{cm}^{2}/(\text{V}\cdot\text{s})$, $T=300$ K,
and $\mu_{c}=0.3$ eV \cite{JAP103-064302,JP19-026222}.
Here, a trilayer graphene is used because it supports plasmonic modes
with higher quality factors
compared with those supported by the monolayer graphene \cite{APL101-111609,JLT32-3597}.

The hybrid modes supported by the hybrid plasmonic
waveguide can be explained using the coupled mode theory (CMT).
In CMT, the mode of an entire
waveguide array is treated as a coupling between modes from single
isolated waveguide channels \cite{waveguide theory}.
\textcolor[rgb]{0.00,0.00,0.00}{Following a similar procedure, the hybrid plasmonic
waveguide shown in Fig. \ref{structure}(a) can be
divided into two isolated waveguide channels
(a dielectric optical waveguide and a plasmonic waveguide)
by increasing the thickness of middle PTFE to infinity,
where the dielectric optical waveguide is composed by air-GaAs-PTFE, and the
plasmonic waveguide is composed by PTFE-graphene-PTFE.
Since the two channels are isolated from each other, the thicknesses of all
PTFE layers in the two channels are infinite.}
For the dielectric optical waveguide, GaAs is used as a high refractive index material
and a single TM optical mode (OM) is supported in the waveguide, where
the dispersion relation and quality factor are
shown by the dashed red lines in Fig. \ref{structure}(b)-(c) and (d), respectively.
Since the dielectric losses are low,
the imaginary part of the propagation constant is small, and the optical mode has
a high quality factor over the chosen frequency range.
While for the plasmonic waveguide,
PTFE is used both as the substrate and superstrate of the trilayer graphene and
a TM plasmonic mode (PM) is supported,
where the dispersion relation and quality factor are
shown by the dashed blue lines. The plasmonic mode has a low quality factor
due to the high Ohmic loss in graphene.
For the hybrid plasmonic waveguide, the optical mode with a high quality factor
and the plasmonic mode with a low quality factor can be coupled
in phase and out of phase, leading to a fundamental mode and a higher order mode,
respectively. As shown by the solid red lines, the fundamental hybrid mode (FHM)
has the largest
$\beta_r$ and the corresponding quality factor is high at higher frequencies
with the decrease of $\beta_{i}$, while the higher order hybrid mode (HOHM)
has the smallest $\beta_r$
and the corresponding quality factor is high at lower frequencies with the decrease of $\beta_{i}$,
as shown by the solid blue lines.

\begin{figure}[ptb]
\centering
\vspace{0.0cm} \includegraphics[width=8.5cm]{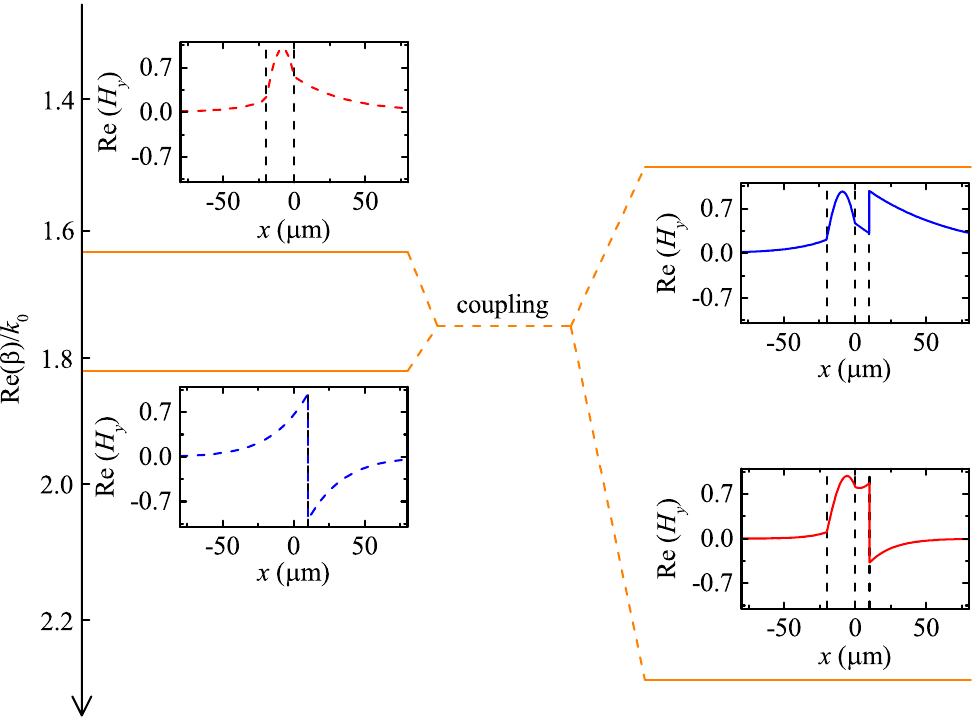} \vspace{-0.0cm}%
\caption{(Color Online)  Propagation constants and corresponding field distributions
of the modes before and after coupling at $f=1.75$ THz. The dashed black lines
indicate the interfaces between different media, \textcolor[rgb]{0.00,0.00,0.00}{and
the dashed yellow lines describe the coupling}. The other parameters are
$t_{1}=20$ $\mu$m, $t_{2}=10$ $\mu$m, $\varepsilon_{1}=1$,
$\varepsilon_{2}=13.06+0.01i$, $\varepsilon_{3}=2.08+0.01i$,
$\mu=$ 10 000 $\text{cm}^{2}/(\text{V}\cdot\text{s})$, $T=300$ K,
and $\mu_{c}=0.3$ eV.}%
\label{1.75}%
\end{figure}

\begin{figure}[ptb]
\centering
\vspace{0.0cm} \includegraphics[width=8.5cm]{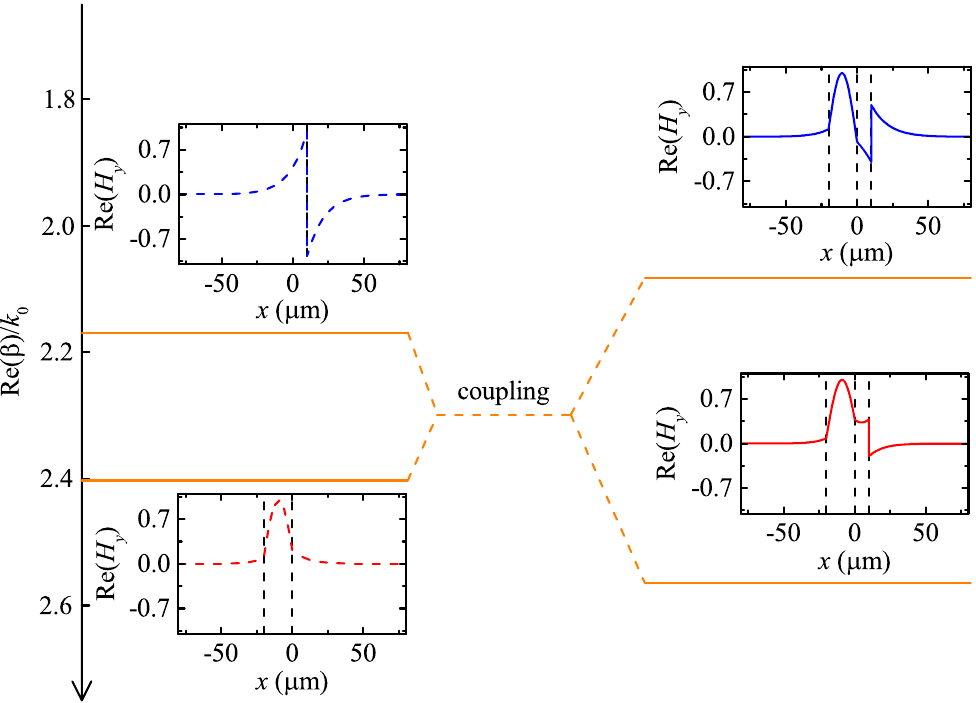} \vspace{-0.0cm}%
\caption{(Color Online) Propagation constants and corresponding field distributions
of the modes before and after coupling at $f=2.50$ THz. The dashed black lines
indicate the interfaces between different media, \textcolor[rgb]{0.00,0.00,0.00}{and
the dashed yellow lines describe the coupling}. The other parameters
are the same as those shown in Fig. \ref{1.75}.}%
\label{2.50}%
\end{figure}

Fig. \ref{1.75} shows the propagation constants and corresponding field distributions
of the modes before and after coupling
at $f=1.75$ THz. Before coupling, the plasmonic waveguide supports
an anti-symmetric TM plasmonic mode with $\beta=1.82$, and the dielectric optical waveguide
supports an asymmetric TM optical mode with $\beta=1.63$, where the quality factors
are $Q=445.4$ and $Q=8.1$, respectively.
At this frequency, the propagation constant of plasmonic mode is larger than that of the optical mode.
After coupling, the fundamental mode has a quality factor of $Q=12.5$ with $\beta=2.30$,
where the optical mode and the plasmonic mode are coupled in phase.
While the higher order mode has a quality factor of $Q=61.4$ with $\beta=1.50$,
where the optical mode and the plasmonic mode are coupled out of phase.
This indicates that the higher order mode is little perturbed by the trilayer graphene
compared with the fundamental mode, and it has a higher quality factor which is favourable
for the implementation of Airy plasmons.

In contrast, Fig. \ref{2.50} shows the propagation constants and corresponding field distributions
of the modes before and after coupling
at $f=2.50$ THz. Before coupling, the plasmonic mode has a quality factor of $Q=8.0$
with $\beta=2.17$, and the optical mode has a quality factor of $Q=907.2$ with $\beta=2.40$.
At this frequency, the propagation constant of plasmonic mode is smaller than that of the optical mode.
After coupling, the fundamental mode has a quality factor of $Q=26.8$ with $\beta=2.57$,
and the higher order mode has a quality factor of $Q=12.0$ with $\beta=2.08$.
In contrast to the hybrid modes at $f=1.75$ THz, the fundamental mode at $f=2.50$ THz
has a higher quality factor compared with the higher order mode.

Clearly, the coupling between the optical mode and the plasmonic mode leads to the
hybrid modes with moderate quality factors. For 1D Airy plasmons, if the waveguide mode
distribution in $x$ direction is chosen as the hybrid mode with a higher quality factor,
we can get a large enough
transverse displacement under the fulfillment of Eq. (\ref{condition_1}).

\subsection{Hybrid Airy plasmons}

\begin{figure}[ptb]
\centering
\vspace{-0.0cm} \includegraphics[width=8.5cm]{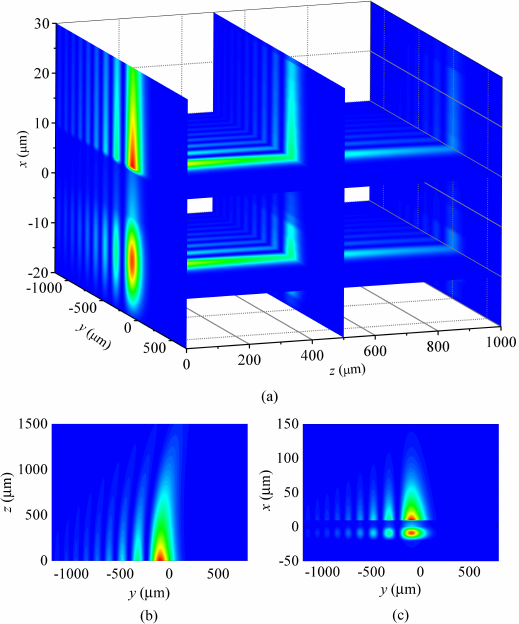} \vspace{-0.0cm}%
\caption{(Color Online) (a) The distributions of
$\left\vert H_{y}\right\vert ^{2}$ for hybrid Airy plasmons on
two $y$-$z$ cut planes with $x=-t_{1}/2$ and $x=t_{2}$, and on three $x$-$y$ cut
planes along the propagation direction with $z=0$ $\mu$m, $z=500$ $\mu$m, and $z=1000$ $\mu$m,
respectively. The distributions at the $y$-$z$ plane with $x=t_{2}$ and the
$x$-$y$ plane with $z=0$ are also shown in (b) and (c), respectively.
The parameters for (a)-(c) are $a=0.1$, $f=1.75$ THz, $y_{0}=100$
$\mu$m, and $\mu_{c}=0.3$ eV.}%
\label{fig4}%
\end{figure}

For Airy plasmons in a hybrid plasmonic waveguide, if their transverse waveguide confinements are
governed by the hybrid modes, these Airy plasmons can be called as \emph{hybrid Airy plasmons}.
In contrast to the Airy beams where the transverse waveguide confinements are
governed by the optical modes \cite{JMO57-341} or plasmonic modes \cite{OL35-2082},
hybrid Airy plasmons are formed due to the coupling between an optical mode and a plasmonic mode.

Without loss of generality, we consider hybrid Airy plasmons
where the transverse waveguide confinements are governed by the higher order hybrid mode
at $f=1.75$ THz,
since it has the highest quality factor within our parameter range.
For simplicity, Fig. \ref{fig4} shows the distributions of $\left\vert H_{y}\right\vert ^{2}$
at different cut planes,
where the scales of $x$, $y$, $z$ axis are different,
and the parameters are $a=0.1$, $y_{0}=100$
$\mu$m,
and $\mu_{c}=0.3$ eV. Note that Eq. (\ref{condition_1}) is valid under the above parameters.
The distribution at the $y$-$z$ plane exhibits
the diffraction-free and self-deflection behaviors of Airy plasmons,
and the distribution at the $x$-$y$ plane
is governed by the higher order hybrid mode in the hybrid plasmonic
waveguide, where
the width of the main lobe of hybrid Airy plasmons is much larger than its height.
It is worth to note that,
although the values of $\left\vert H_{y}\right\vert ^{2}$
at two $y$-$z$ planes are nearly
equal, the graphene-based plasmonic waveguide channel carries more energy if we use
the standard definition of energy flux density $\left\langle S\right\rangle _{z}=\frac{1}{2}\text{Re}\left(
E_{x}H_{y}^{\ast }\right) $.

As shown in Fig. \ref{fig4}(b), due to the high quality factor of the hybrid mode,
the energy attenuation experienced
by Airy plasmons is relatively small and the propagation length is comparably large,
although the dissipative loss exists both in dielectric media and in trilayer graphene.
Under the chosen parameters, the propagation length is
$L_{n}=537.5$ $\mu$m ($L_{a}=558.9$ $\mu$m), and the corresponding transverse
displacement at the propagation length is $\Delta y_{n}=24.0$ $\mu$m ($\Delta
y_{a}=26.0$ $\mu$m), where the analytical results are nearly equal to the
numerical results. The large transverse displacement
is favourable for the detection and measurement of Airy plasmons experimentally.
It is worth to note that, if the transverse waveguide confinement of Airy plasmons
is governed by the plasmonic mode shown in Fig. \ref{1.75} with $a=0.1$ and
$y_{0}=100$ $\mu$m, the propagation length is
$L_{n}=0.6$ $\mu$m ($L_{a}=0.6$ $\mu$m) and the corresponding transverse
displacement at the propagation length is $\Delta y_{n}=1.0$ $\mu$m ($\Delta
y_{a}=0.2$ $\mu$m). Our hybrid Airy plasmons show a better performance
due to the moderate quality factor of the hybrid mode.

\begin{figure*}[ptb]
\centering
\vspace{-0.0cm} \includegraphics[width=14cm]{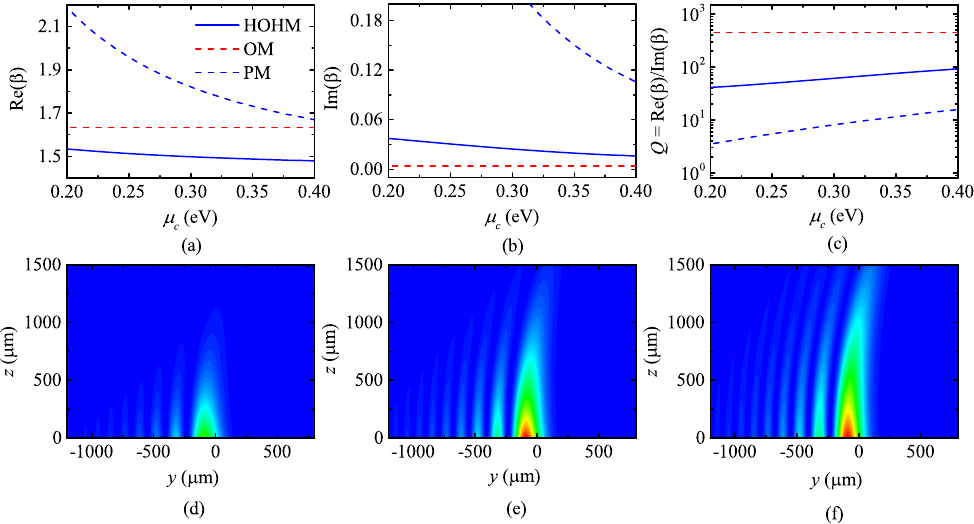} \vspace{-0.0cm}%
\caption{(Color Online) (a)-(c) The real and imaginary
parts of the propagation constants, and the quality factors of the higher order
hybrid mode (HOHM) versus the chemical potential of trilayer graphene, respectively.
For comparison, the curves for the optical mode (OM) supported by air-GaAs-PTFE and plasmonic
mode (PM) supported by PTFE-graphene-PTFE are also plotted in dashed red and blue,
respectively, and the curves for the fundamental hybrid mode are omitted.
(d)-(e) The distributions of $\left\vert H_{y}\right\vert ^{2}$ at the $y$-$z$ plane with
$x = t_{2}$ for $\mu_{c}=0.2$ eV, $\mu_{c}=0.3$ eV, and $\mu_{c}=0.4$ eV, respectively.
The other parameters are $a=0.1$, $f=1.75$ THz, and $y_{0}=100$ $\mu$m.}%
\label{fig5}%
\end{figure*}

\begin{figure}[ptb]
\centering
\vspace{0.0cm} \includegraphics[width=8.5cm]{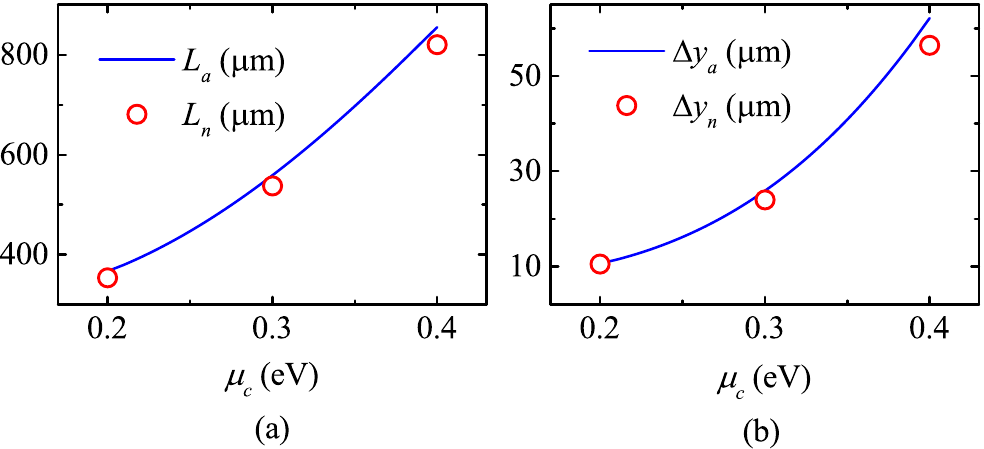} \vspace{-0.0cm}%
\caption{(Color Online) (a) Analytical propagation length $L_{a}$
and (b) transverse displacement $\Delta y_{a}$ of Airy plasmons
versus the chemical potential of trilayer graphene.
The numerical propagation length $L_{n}$ and transverse displacement
$\Delta y_{n}$ for $\mu_{c}=0.2$ eV, $\mu_{c}=0.3$ eV, and $\mu_{c}=0.4$ eV
are also plotted for comparison.
The other parameters are $a=0.1$, $f=1.75$ THz, and $y_{0}=100$ $\mu$m.}%
\label{fig6}%
\end{figure}

For the effective trapping, transporting, and sorting of micro-objects,
Airy plasmons with dynamically steerable trajectories are required.
Since the self-deflection behavior is related with the propagation
constant of the hybrid mode, we can tune the chemical
potential of trilayer graphene dynamically by applying a gate voltage \cite{science332-1291}.
As shown in Fig. \ref{fig5} (a)-(c), for the higher order hybrid mode (HOHM),
both the propagation constant
and quality factor change effectively if the chemical potential is
tuned from $0.2$ eV to $0.4$ eV (the curves for the fundamental hybrid mode
are omitted), where the parameters are
$a=0.1$, $f=1.75$ THz, and $y_{0}=100$ $\mu$m to insure the validity of Eq.
(\ref{condition_1}).
Clearly, when the
chemical potential is large, the imaginary part of the propagation constant
$\beta_{i}$ is small and the corresponding propagation length is large. The
increases of the propagation length and quality factor lead to a larger
transverse displacement of Airy plasmons, as shown by the distributions
of $\left\vert H_{y}\right\vert ^{2}$
at the $y$-$z$ plane with $x = t_2$ for $\mu_{c}=0.2$ eV, $\mu_{c}=0.3$ eV,
and $\mu_{c}=0.4$ eV in Fig. \ref{fig5} (a)-(c), respectively.
Thus the transverse displacement is dynamically steerable by changing the chemical
potential of graphene (See Supporting Information).

To quantitatively show the steerable trajectories by tuning the chemical potential
of trilayer graphene, we plot the numerical propagation
lengths $L_n$ and transverse displacements $\Delta y_n$ for $\mu_{c}=0.2$ eV, $\mu_{c}=0.3$ eV,
and $\mu_{c}=0.4$ eV as well as the analytical counterparts in Fig. \ref{fig6}.
The consistency between the analytical and numerical results not only confirms
the validity of the paraxial approximation which insures the non-diffracting behavior
of Airy plasmons, but also demonstrates the feasibility of dynamically steerable
trajectories of hybrid Airy plasmons.

\textcolor[rgb]{0.00,0.00,0.00}{Finally, we would like to note that,
our hybrid Airy plasmons can be realized experimentally by considering both the
waveguide fabrication and Airy plasmons excitation. For the waveguide fabrication,
all the materials including GaAs, PTFE, and graphene are widely used in THz waveguides.
Graphene can be grown well by a CVD reaction between CH$_{\text{4}}$ and H$_{\text{2}}$ on copper substrate,
which can be etched away using acid solution. The suspended graphene can subsequently
be transferred to arbitrary substrate. Recently, we have designed a graphene/GaAs
van der Waals heterostructure solar cell \cite{NE16-310} and studied the electric power generation
from a single moving droplet on graphene/PTFE \cite{ACSNano10-7297}. The wet transfer technique that used
in the two experiments can be directly applied to fabricate the hybrid graphene-based plasmonic waveguide.
While for the Airy plasmons excitation, it includes two steps:
the generation of surface plasmon polaritons (SPPs) and the generation of Airy plasmons.
The two steps can be accomplished simultaneously by a delicately designed grating \cite{PRL107-116802,OL38-1443}.
Alternatively,
a laser beam can be coupled into SPPs by grating, and Airy plasmons can be generated
by a nonperiodically arranged nanocave array \cite{PRL107-126804}.}

\section{Conclusion}

In conclusion, we reveal the hybrid Airy plasmons for the first time
by taking a hybrid graphene-based plasmonic waveguide in the THz domain as an example.
Due to the moderate quality factors of the hybrid modes,
hybrid Airy plasmons where the transverse waveguide confinements are governed by
the hybrid modes can have large propagation lengths and effective transverse deflections,
and they are promising to solve the long standing problems
for the experimental observation and implementation of Airy plasmons.
Meanwhile, the propagation trajectories of hybrid
Airy plasmons are dynamically steerable by changing the chemical potential of
trilayer graphene.
This interesting finding may lead to the flexible optical micromanipulation
in flatland plasmonic devices and chip scale signal processing,
and the hybrid Airy plasmons may promote the further
discovery of non-diffracting beams with the emerging developments of optical tweezers and
tractor beams.

\section{Methods}

\subsection{Paraxial Wave Equation}

Airy beams propagating in a planar waveguide can be described by the 1D
paraxial wave equation which is similar with the Schr\"{o}dinger equation
in quantum mechanics. For simplicity, we assume that Airy beams are
perturbations of the waveguide modes, and they behave as quasi-TM or quasi-TE
modes approximately. The Helmholtz equation is
\begin{equation}
\nabla^{2}\Phi+k_{0}^{2}\varepsilon\left(  x\right)  \Phi=0,
\label{Helmholtz equation}%
\end{equation}
where $\varepsilon\left(  x\right)  $ is the distribution of relative
permittivity, $\Phi=H_{y}(x,y,z)$ for the quasi-TM Airy beams, and $\Phi
=E_{y}(x,y,z)$ for the quasi-TE Airy beams. The scalar field $\Phi\left(
x,y,z\right)  $ can be expressed as a functional dependence of the form
\cite{JMO57-341}
\begin{equation}
\Phi\left(  x,y,z\right)  =\psi\left(  y,z\right)  \Phi_{0}\left(  x\right)
\exp\left(  i\beta z\right)  , \label{field}%
\end{equation}
where the dimensionless scalar function $\psi\left(  y,z\right)  $ is
dependent on both the transverse direction $y$ and the propagation direction
$z$, the scalar function $\Phi_{0}\left(  x\right)  $ is only dependent on the
transverse direction $x$, and $\beta$ is a parameter that is related to the
$z$ component of the wavevector. Substituting Eq. (\ref{field}) into Eq.
(\ref{Helmholtz equation}), multiplying the result by $\Phi_{0}^{\ast}\left(
x\right)  $, and integrating over $x$ direction yields a scalar wave equation
\begin{align}
&  \left[  2i\beta\frac{\partial\psi}{\partial z}+\frac{\partial^{2}\psi
}{\partial y^{2}}\right]  +\left[  \frac{I_{2}}{I_{0}}+k_{0}^{2}%
\varepsilon\left(  x\right)  -\beta^{2}\right]  \psi=0, \label{wave equation}%
\end{align}
where $I_{0}=\int_{-\infty}^{+\infty}\left\vert \Phi_{0}\left(  x\right)
\right\vert ^{2}dx$, $I_{2}=\int_{-\infty}^{+\infty}\Phi_{0}^{\prime\prime
}\left(  x\right)  \Phi_{0}^{\ast}\left(  x\right)  dx$, and the term
$\partial^{2}\psi/\partial z^{2}$ is neglected by employing the paraxial
approximation \cite{OL35-2082,OL32-674}. Note the
functional dependence between $\psi$ and $x$ is neglected for step index
waveguides or graded index waveguides with $\nabla\varepsilon\approx0$.
For the slowly varying amplitude, the
scalar function is expressed as $\psi\left(  y,z\right)  =\phi\left(
y,z\right)  \exp\left\{  i\left[  I_{2}/I_{0}+k_{0}^{2}\varepsilon\left(
x\right)  -\beta^{2}\right]  /\left(  2\beta\right)  z\right\}  $, which leads
to the one-dimensional paraxial wave equation $i\phi_{z}\left(  y,z\right)
+\left(  1/2\beta\right)  \phi_{yy}\left(  x,z\right)  =0$. If the
function $\Phi_{0}\left(  x\right)  $ is chosen as the magnetic (electric)
field distribution of TM (TE) waveguide mode, and the parameter $\beta
=\beta_{r}+i\beta_{i}$ is the corresponding propagation constant, $I_{2}%
/I_{0}+k_{0}^{2}\varepsilon\left(  x\right)  -\beta^{2}=0$ and the
paraxial wave equation for the amplitude $\psi$ is%
\begin{equation}
i\frac{\partial\psi\left(  s,\xi\right)  }{\partial\xi}+\frac{\partial^{2}%
\psi\left(  s,\xi\right)  }{\partial s^{2}}=0, \label{schrodinger}%
\end{equation}
where $s=y/y_{0}$ is the dimensionless transverse coordinate, $\xi=z/2\beta
y_{0}^{2}$ is the dimensionless complex propagation distance, and $y_{0}$ is
an arbitrary transverse scale.

\subsection{Dispersion relation for hybrid modes}

Dispersion relation for hybrid modes in the hybrid graphene-based planar
plasmonic waveguide can be calculated by standard waveguide theory \cite{waveguide theory},
where the field
components in each region are derived from Helmholtz equation, and the
propagation constant is determined by imposing appropriate boundary conditions.
For the structure shown in Fig. 1, the magnetic field
distribution can be written as follows%
\begin{equation}
H_{y}\left(  x\right)  =\left\{
\begin{array}
[c]{ll}%
Ae^{k_{1}x}e^{i\beta z}, & x\leq-t_{1},\\
B\sin\left(  k_{2}x\right)  e^{i\beta z}+C\cos\left(  k_{2}x\right)  e^{i\beta
z}, & -t_{1}<x\leq0,\\
De^{k_{3}x}e^{i\beta z}+Ee^{-k_{3}x}e^{i\beta z}, & 0<x\leq t_{2},\\
Fe^{-k_{3}x}e^{i\beta z}, & x>t_{2},
\end{array}
\right.  \label{mode}%
\end{equation}
where $k_{1}=\sqrt{\beta^{2}-k_{0}^{2}\varepsilon_{1}}$, $k_{2}=\sqrt
{k_{0}^{2}\varepsilon_{2}-\beta^{2}}$, $k_{3}=\sqrt{\beta^{2}-k_{0}%
^{2}\varepsilon_{3}}$, $k_{0}=2\pi f/c$ is the wavenumber in free space, $f$
is the frequency, and $\varepsilon_{1}$, $\varepsilon_{2}$, and $\varepsilon
_{3}$ are relative permittivities of air, GaAs, and PTFE, respectively.
Considering the boundary conditions, the dispersion relation is%
\begin{equation}
e^{-2k_{3}t_{2}}=\frac{\alpha_{+}\left[  \beta_{-}\tan\left(  k_{2}%
t_{1}\right)  +\gamma_{+}\right]  }{\alpha_{-}\left[  \beta_{+}\tan\left(
k_{2}t_{1}\right)  -\gamma_{-}\right]  }, \label{dispersion relation}%
\end{equation}
where $\alpha_{\pm}=\left(  1+i\sigma k_{3}/\omega\varepsilon_{0}%
\varepsilon_{3}\right)  \left(  k_{3}/\varepsilon_{3}\right)  \pm\left(
k_{3}/\varepsilon_{3}\right)  $, $\beta_{\pm}=\left(  k_{1}/\varepsilon
_{1}\right)  \left(  k_{3}/\varepsilon_{3}\right)  \pm\left(  k_{2}%
/\varepsilon_{2}\right)  ^{2}$,
$\gamma_{\pm}=\left(  k_{2}/\varepsilon
_{2}\right)  \left[  \left(  k_{1}/\varepsilon_{1}\right)  \pm\left(
k_{3}/\varepsilon_{3}\right)  \right]  $, and $\sigma$ is the surface
conductivity of trilayer graphene.

The hybrid modes are formed by the coupling between the optical modes in
dielectric waveguide and the plasmonic modes in graphene-based plasmonic
waveguide. The dispersion relation for the optical modes in dielectric
waveguide is
\begin{equation}
\tan\left(  k_{2}t_{1}\right)  =-\frac{\gamma_{+}}{\beta_{-}},
\label{dielectric dispersion}%
\end{equation}
which can be obtained by setting $\sigma=0$ in Eq. (\ref{dispersion relation}%
). While for the plasmonic modes in graphene-based plasmonic waveguide
\cite{science332-1291}, the dispersion relation is
\begin{equation}
\beta=k_{0}\sqrt{\varepsilon_{3}-\left(  \frac{2\varepsilon_{3}}{\sigma
\eta_{0}}\right)  ^{2}}, \label{plasmonic dispersion}%
\end{equation}
where $\eta_{0}=\sqrt{\mu_{0}/\varepsilon_{0}}$ is the wave impendence in free space.

The surface conductivity of monolayer graphene can be calculated by Kubo
formula $\sigma_{g}\left(  \omega,\mu_{c},\Gamma,T\right)  =\sigma_{\text{intra}%
}+\sigma_{\text{inter}}$, where
\begin{equation}
\sigma_{\text{intra}}=\frac{ie^{2}k_{B}T}{\pi\hbar^{2}\left(  \omega
+i\tau^{-1}\right)  }\left[  \frac{\mu_{c}}{k_{B}T}+2\ln\left(  e^{-\mu
_{c}/k_{B}T}+1\right)  \right]  \label{intra}%
\end{equation}
is due to intraband contribution,
\begin{equation}
\sigma_{\text{inter}}=\frac{ie^{2}\left(  \omega+i\tau^{-1}\right)  }{\pi
\hbar^{2}}\int_{0}^{\infty}\frac{f_{d}\left(  -\varepsilon\right)
-f_{d}\left(  \varepsilon\right)  }{\left(  \omega+i\tau^{-1}\right)
^{2}-4\left(  \varepsilon/\hbar\right)  ^{2}}d\varepsilon\label{inter}%
\end{equation}
is due to interband contribution \cite{JAP103-064302,JP19-026222},
and the nonlocal effects are neglected \cite{PRB80-245435}. In the
above formula, $-e$ is the charge of an electron, $\hbar=h/2\pi$ is the
reduced Plank's constant, $T$ is the temperature, $\mu_{c}$ is the chemical
potential, $\tau=\mu\mu_{c}/ev_{F}^{2}$ is the carrier relaxation time, $\mu$
is the carrier mobility which ranges from 1 000 $\text{cm}^{2}/(\text{V}%
\cdot\text{s})$ to 230 000 $\text{cm}^{2}/(\text{V}\cdot\text{s})$
\cite{ACSnano}, $v_{F}=c/300$ is the Fermi velocity, $f_{d}\left(
\varepsilon\right)  =1/\left[  e^{\left(  \varepsilon-\mu_{c}\right)  /k_{B}%
T}+1\right]  $ is the Fermi-Dirac distribution, and $k_{B}$ is the Boltzmann's
constant. In this paper, we use a moderate mobility of $\mu=$ 10 000
$\text{cm}^{2}/(\text{V}\cdot\text{s})$, $T=300$ K, and the chemical potential
$\mu_{c}$ is tuned from $0.2$ eV to $0.4$ eV.

\section{acknowledgement}

We are grateful to Mr. Huikai Zhong for his valuable discussion.
This work was sponsored by the National Natural Science Foundation of China
under Grants No. 61322501, No. 61574127, and No. 61275183, the Top-Notch Young
Talents Program of China, the Program for New Century Excellent Talents
(NCET-12-0489) in University, the Fundamental Research Funds for the Central
Universities, and the Innovation Joint Research Center for
Cyber-Physical-Society System. X. Lin acknowledges the support from Chinese
Scholarship Council (CSC No. 201506320075).


\begin{thebibliography}{99}                                                                                               %

\bibitem {AJP}M. V. Berry and N. L. Balazs, \emph{Nonspreading wave packets},
Am. J. Phys. \textbf{47}, 264 (1979)

\bibitem {OL32-979}G. A. Siviloglou and D. N. Christodoulides,
\emph{Accelerating finite energy Airy beams}, Opt. Lett. \textbf{32}, 979 (2007).

\bibitem {LPR4-529}M. Mazilu, D. J. Stevenson, F. G.-Moore, and K. Dholakia,
\emph{Light beats the spread: ``non-diffracting'' beams}, Laser Photonics Rev.
\textbf{4}, 529 (2010).

\bibitem {OL32-2447}I. M. Besieris and A. M. Shaarawi, \emph{A note on an
accelerating finite energy Airy beam}, Opt. Lett. \textbf{32}, 2447 (2007).

\bibitem {PRL99-213901}G. A. Siviloglou, J. Broky, A. Dogariu, and D. N.
Christodoulides, \emph{Observation of accelerating Airy beams}, Phys. Rev.
Lett. \textbf{99}, 213901 (2007).

\bibitem {OE16-12880}J. Broky, G. A. Siviloglou, A. Dogariu, and D. N.
Christodoulides, \emph{Self-healing properties of optical Airy beams}, Opt.
Exp. \textbf{16}, 12880 (2008).

\bibitem {OL35-4045}N. K. Efremidis and D. N. Christodoulides, \emph{Abruptly
autofocusing waves}, Opt. Lett. \textbf{35}, 4045 (2010).

\bibitem {OL36-1842}D. G. Papazoglou, N. K. Efremidis, D. N. Christodoulides,
and S. Tzortzakis, \emph{Observation of abruptly autofocusing waves}, Opt.
Lett. \textbf{36}, 1842 (2011).

\bibitem {nphoton2-675}J. Baumgartl, M. Mazilu, and K. Dholakia,
\emph{Optically mediated particle clearing using Airy wavepackets}, Nat.
Photon. \textbf{2}, 675 (2008).

\bibitem {OL36-2883}P. Zhang, J. Prakash, Z. Zhang, M. S. Mills, N. K.
Efremidis, D. N. Christodoulides, and Z. Chen, \emph{Trapping and guiding
microparticles with morphing autofocusing Airy beams}, Opt. Lett. \textbf{36},
2883 (2011).

\bibitem {AO50-43}Z. Zheng, B.-F. Zhang, H. Chen, J. Ding, and H.-T. Wang,
\emph{Optical trapping with focused Airy beams}, Appl. Opt. \textbf{50}, 43 (2011).

\bibitem {OL40-5686}Y. Liang, Y. Hu, D. Song, C. Lou, X. Zhang, Z. Chen, and
J. Xu, \emph{Image signal transmission with Airy beams}, Opt. Lett.
\textbf{40}, 5686 (2015).

\bibitem {APL102-101101}P. Rose, F. Diebel, M. Boguslawski, and C. Denz,
\emph{Airy beam induced optical routing}, Appl. Phys. Lett. \textbf{102},
101101 (2013).

\bibitem {OL39-5997}N. Wiersma, N. Marsal, M. Sciamanna, and D. Wolfersberger,
\emph{All-optical interconnects using Airy beams}, Opt. Lett. \textbf{39},
5997 (2014).

\bibitem {nphoton4-103}A. Chong, W. H. Renninger, D. N. Christodoulides, and
F. W. Wise, \emph{Airy-Bessel wave packets as versatile linear light bullets},
Nat. Photon. \textbf{4}, 103 (2010).

\bibitem {OE16-10303}P. Saari, \emph{Laterally accelerating Airy pulses}, Opt.
Exp. \textbf{16}, 10303 (2008).

\bibitem {OE19-2286}K.-Y. Kim, C.-Y. Hwang, and B. Lee, \emph{Slow
non-dispersing wavepackets}, Opt. Exp. \textbf{19}, 2286 (2011).

\bibitem {PRL104-197203}T. Schneider, A. A. Serga, A. V. Chumak, C. W.
Sandweg, S. Trudel, S. Wolff, M. P. Kostylev, V. S. Tiberkevich, A. N. Slavin,
and B. Hillebrands, \emph{Nondiffractive subwavelength wave beams in a medium
with externally controlled anisotropy}, Phys. Rev. Lett. \textbf{104}, 197203 (2010).



\bibitem {PRL115-034501}S. Fu, Y. Tsur, J. Zhou, L. Shemer, and A. Arie,
\emph{Propagation Dynamics of Airy Water-Wave Pulses}, Phys. Rev. Lett.
\textbf{115}, 034501 (2015).

\bibitem {nature}N. V.-Bloch, Y. Lereah, Y. Lilach, A. Gover, and A. Arie,
\emph{Generation of electron Airy beams}, Nature \textbf{494}, 331 (2013).

\bibitem {PRA87-043637}N. K. Efremidis, and V. Paltoglou, \emph{Accelerating
and abruptly autofocusing matter waves}, Phys. Rev. A \textbf{87}, 043637 (2013).

\bibitem {OL35-2082}A. Salandrino, and D. N. Christodoulides, \emph{Airy
plasmon: a nondiffracting surface wave}, Opt. Lett. \textbf{35}, 2082 (2010).

\bibitem {ieeepj}Y. Yang, H. T. Dai, B. F. Zhu, and X. W. Sun, \emph{Dynamic
Control of the Airy Plasmons in a Graphene Platform}, IEEE Photonics J.
\textbf{6}, 4801207 (2014).

\bibitem {PRL107-116802}A. Minovich, A. E. Klein, N. Janunts, T. Pertsch, D.
N. Neshev, and Y. S. Kivshar, \emph{Generation and Near-Field Imaging of Airy
Surface Plasmons}, Phys. Rev. Lett. \textbf{107}, 116802 (2011).

\bibitem {PRL107-126804}L. Li, T. Li, S. M. Wang, C. Zhang, and S. N. Zhu,
\emph{Plasmonic Airy Beam Generated by In-Plane Diffraction}, Phys. Rev. Lett.
\textbf{107}, 126804 (2011).

\bibitem {OL34-3430}A. V. Novitsky, and D. V. Novitsky, \emph{Nonparaxial Airy
beams: role of evanescent waves}, Opt. Lett. \textbf{34}, 3430 (2009).

\bibitem {OL36-3191}P. Zhang, S. Wang, Y. Liu, X. Yin, C. Lu, Z. Chen, and X.
Zhang, \emph{Plasmonic Airy beams with dynamically controlled trajectories},
Opt. Lett. \textbf{36}, 3191 (2011).

\bibitem {OL36-1164}W. Liu, D. N. Neshev, I. V. Shadrivov, A. E.
Miroshnichenko, and Y. S. Kivshar, \emph{Plasmonic Airy beam manipulation in
linear optical potentials}, Opt. Lett. \textbf{36}, 1164 (2011).

\bibitem {OL38-1443}F. Bleckmann, A. Minovich, J. Frohnhaus, D. N. Neshev, and
S. Linden, \emph{Manipulation of Airy surface plasmon beams}, Opt. Lett.
\textbf{38}, 1443 (2013).

\bibitem {LPR8-221}A. E. Minovich, A. E. Klein, D. N. Neshev, T. Pertsch, Y.
S. Kivshar, and D. N. Christodoulides, \emph{Airy plasmons: non-diffracting
optical surface waves}, Laser Photon. Rev. \textbf{8}, 221 (2014).

\bibitem{nphoton2-496}R. F. Oulton, V. J. Sorger, D. A. Genov, D. F. P. Pile, and X. Zhang,
\emph{A hybrid plasmonic waveguide for subwavelength confinement and long range
propagation}, Nat. Photon. \textbf{2}, 496 (2008).

\bibitem {JLT32-3597}X. Zhou, T. Zhang, L. Chen, W. Hong, and X. Li, \emph{A
Graphene-Based Hybrid Plasmonic Waveguide With Ultra-Deep Subwavelength
Confinement}, J. Lighwave Techno. \textbf{32}, 3597 (2014).

\bibitem {Airy-function}M. Abramowitz, and I. A. Stegun, \emph{Handbook of
Mathematical Functions} (Dover, 1972).

\bibitem{acsphotonics3-737}A. R. Davoyan and N. Engheta,
\emph{Salient Features of Deeply Subwavelength Guiding of Terahertz
Radiation in Graphene-Coated Fibers}, ACS Photonics \textbf{3}, 737 (2016).

\bibitem{Nanotechnology}R. Li, X. Lin, S. Lin, X. Liu, and
Hongsheng Chen, \emph{Atomically thin spherical shell-shaped
superscatterers based on a Bohr model}, Nanotechnology \textbf{26},
505201 (2015).

\bibitem{IEEEJSQE}R. Li, B. Zheng, X. Lin, R. Hao, S. Lin,
W. Yin, E. Li, and H. Chen, \emph{Design of Ultracompact Graphene-Based
Superscatterers}, IEEE J. Sel. Top. Quant. \textbf{23}, 4600208 (2017).

\bibitem {science332-1291}A. Vakil, and N. Engheta, \emph{Transformation
Optics Using Graphene}, Science \textbf{332}, 1291 (2011).

\bibitem {nmat1-26}B. Ferguson and X.-C. Zhang,
\emph{Materials for terahertz science and technology},
Nat. Mat. \textbf{1}, 26 (2002).

\bibitem{optical constants}D. Palik, \emph{Handbook of
Optical Constants of Solids} (Academic Press, 1998).

\bibitem{JKPS}Y.-S. Jin, G.-J. Kim, and S.-G. Jeon, \emph{Terahertz Dielectric Properties of Polymers},
J. Korean Phys. Soc. \textbf{49}, 513 (2006).

\bibitem {NL7-2711}C. Casiraghi, A. Hartschuh, E. Lidorikis, H. Qian, H.
Harutyunyan, T. Gokus, K. S. Novoselov, and A. C. Ferrari, \emph{Rayleigh
Imaging of Graphene and Graphene Layers}, Nano Lett. \textbf{7}, 2711 (2007).

\bibitem {nnano7-330}H. Yan, X. Li, B. Chandra, G. Tulevski, Y. Wu, M.
Freitag, W. Zhu, P. Avouris, and F. Xia, \emph{Tunable infrared plasmonic
devices using graphene/insulator stacks}, Nat, Nanotech. \textbf{7}, 330 (2012).

\bibitem {LPR8-291}D. A. Smirnova, I. V. Shadrivov, A. I. Smirnov, and Y. S.
Kivshar, \emph{Dissipative plasmon-solitons in multilayer graphene}, Laser
Photonics Rev. \textbf{8}, 291 (2014).

\bibitem {JAP103-064302}G. W. Hanson, \emph{Dyadic Green's functions and
guided surface waves for a surface conductivity model of graphene}, J. Appl.
Phys. \textbf{103}, 064302 (2008).

\bibitem {JP19-026222}V. P. Gusynin, S. G. Sharapov, and J. P. Carbotte,
\emph{Magneto-optical conductivity in graphene}, J. Phys. \textbf{19}, 026222 (2007).

\bibitem {APL101-111609}C. H. Gan, \emph{Analysis of surface plasmon
excitation at terahertz frequencies with highly doped graphene sheets via
attenuated total reflection}, Appl. Phys. Lett. \textbf{101}, 111609 (2012).

\bibitem {waveguide theory}K. Okamoto, \emph{Fundamentals of Optical
Waveguides} (Elsevier, 2006).

\bibitem {JMO57-341}V. Lakshminarayanana, and K. Thyagarajan,
\emph{Non-diffracting Airy beams in planar optical waveguides: a convenient
method for visualization}, J. Mod. Opt. \textbf{57}, 341 (2010).

\bibitem{NE16-310}\textcolor[rgb]{0.00,0.00,0.00}{X. Li, W. Chen, S. Zhang, Z. Wu, P. Wang, Z. Xu, H. Chen, W. Yin,
H. Zhong, S. Lin, \emph{18.5\% efficient graphene/GaAs van der Waals heterostructure solar cell},
Nano Energy \textbf{16}, 310 (2015).}

\bibitem{ACSNano10-7297}\textcolor[rgb]{0.00,0.00,0.00}{S. S. Kwak, S. Lin, J. H. Lee, H. Ryu, T. Y. Kim, H. Zhong,
H. Chen, and S.-W. Kim, \emph{Triboelectrification-Induced Large Electric
Power Generation from a Single Moving Droplet on Graphene/Polytetrafluoroethylene},
ACS Nano \textbf{10}, 7297 (2016).}

\bibitem {OL32-674}E. Feigenbaum, and M. Orenstein, \emph{Plasmon-soliton},
Opt. Lett. \textbf{32}, 674 (2007).

\bibitem{PRB80-245435}M. Jablan, H. Buljan, and M. Solja\v{c}i\'{c}, \emph{Plasmonics in graphene
at infrared frequencies}, Phys. Rev. B \textbf{80}, 245435 (2009).

\bibitem {ACSnano}W. Gao, J. Shu, C. Qiu, and Q. Xu, \emph{Excitation of
Plasmonic Waves in Graphene by Guided-Mode Resonances}, ACS Nano \textbf{6},
7806 (2012).




\end{thebibliography}
\end{document}